\documentclass[aps,prd,reprint,nofootinbib]{revtex4-2}

\usepackage{amsmath,amssymb,graphicx,bm,microtype,mathrsfs,booktabs}
\usepackage{microtype}
\usepackage{graphicx}

\usepackage{physics}   
\usepackage{tikz}
\usepackage{tikz-cd}

\usepackage[colorlinks=true,allcolors=blue]{hyperref}
\usepackage[nameinlink,capitalise]{cleveref}

\usepackage{enumitem}
\setlist[enumerate]{label=(\arabic*),leftmargin=2.2em,itemsep=0pt,topsep=2pt}

\crefname{equation}{Eq.}{Eqs.}
\Crefname{equation}{Equation}{Equations}
\crefname{figure}{Fig.}{Figs.}
\Crefname{figure}{Figure}{Figures}

\newcommand{\Prob}{\mathbb P}

\begin{document}

\preprint{}

\title{Quantum Selection of Classical Histories}

\author{Omer Guleryuz}
\email{omerguleryuz@itu.edu.tr}

\affiliation{Department of Physics, Istanbul Technical University, Maslak 34469 Istanbul, Türkiye}

\date{\today}

\begin{abstract}
Can a finite quantum episode populate a classical future that the same prepared state never reaches deterministically? We show that it can, using inflation as a cosmological laboratory. In our supergravity realization, two classical continuations already exist, but the prepared state reaches only one without noise. A stable entropy direction temporarily becomes light without a tachyonic instability, allowing mode-derived phase-space fluctuations to access the competing continuation. Finite-horizon first passage assigns it a \(0.33\)--\(0.40\) conditional weight across three nearby CMB-normalized realizations, while suppressing the stochastic source restores its exact deterministic zero. The entropy sector then becomes heavy again, and the selected histories continue classically. A finite quantum episode can thus leave a lasting redistribution among classical futures.
\end{abstract}

\maketitle

\section{Introduction}
Classical equations assign a unique trajectory once complete phase-space data are fixed. Quantum fluctuations need not alter those equations, create a new solution, or compromise that deterministic uniqueness. They can instead redistribute finite-history weight among classical continuations that already exist in the solution space. A particularly sharp possibility is transient: a history prepared on one classical continuation encounters a finite quantum window in which a second continuation acquires nonzero weight, and then returns to a regime in which both outgoing histories evolve deterministically. Fluctuations create no new classical future; they assign nonzero weight to a future whose weight is exactly zero in the same prepared state without noise. Here, a ``history weight'' denotes the finite-horizon first-passage probability for this dynamical problem, not a global cosmological measure. Inflation provides a controlled cosmological setting in which this quantum-to-classical sequence can be followed explicitly.

In multifield inflation, this regime lies between several familiar limits. A permanently heavy entropy mode may be integrated out \cite{Achucarro:2012sm}, whereas a tachyonic waterfall or geometrical destabilization opens new evolution through an instability of the background \cite{Linde:1993cn,Martin:2011ib,Renaux-Petel:2015mga,Gong:2022tfu}. Stochastic or eternal inflation instead keeps fluctuations dynamically important over an extended interval \cite{Starobinsky:1986fx,Starobinsky:1994bd}. Multiple rolling channels can already coexist classically \cite{Li:2009sp,Wang:2010rs,Duplessis:2012nb}, but their existence does not tell us whether a fixed prepared state can populate each of them. Stationary stochastic double-well equilibria address a different question \cite{Markkanen:2020bfc,Camargo-Molina:2022ord}. We ask instead whether a stable transverse direction can become light only long enough to alter the accessibility of two pre-existing continuations, and then become heavy again.

\begin{figure}[htb!]
 \centering
 \includegraphics[width=\columnwidth]{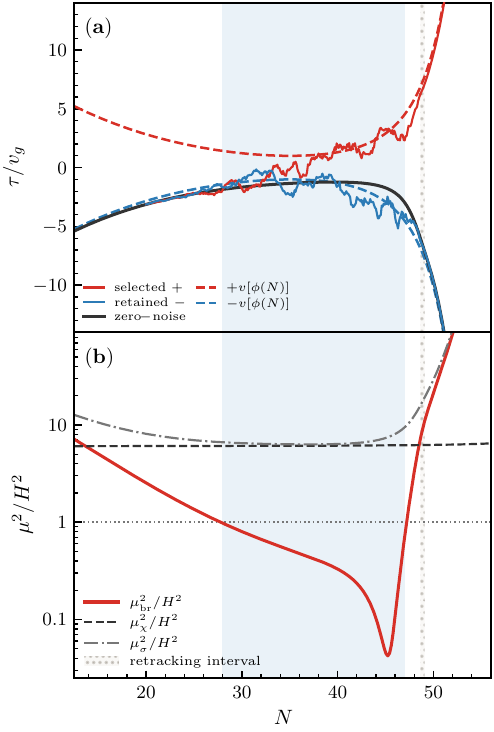}
\caption{\textbf{Quantum selection and classical recovery} for
\(\alpha=1\). (a) Selected histories, the zero-noise trajectory, and the moving channels; blue shading marks the light interval. (b) Entropy diagnostics; the gray band separates heavy-sector recovery from subsequent branch tracking.}
 \label{fig:dynamics}
\end{figure}

We realize this sequence explicitly in \(\mathcal N=1\) supergravity. A holomorphic branch equation defines two reflection-related classical continuations, while the nilpotent inverse K\"ahler metric supplies their rank-one restoring sector. The branches first narrow and then reopen; because the rolling solution follows their moving centers with a finite response time, the delayed response after the geometric turnaround deepens a transient softening without ever making the relevant entropy mode tachyonic. We solve the complete coupled scalar modes, retain their field--momentum covariance, propagate the resulting nonlinear stochastic phase-space dynamics, and define selection by competing first passage into recovered classical channels. Recovery of the heavy regime and subsequent branch tracking then test whether the selected histories continue without stochastic support.

Figure~\ref{fig:dynamics} illustrates what this sequence means in terms of dynamics. The prepared zero-noise history remains on the negative continuation, while the finite light interval populates both outgoing channels. The blue band marks the only interval in which the branch entropy direction is light. Its end does not by itself complete the recovery: the restoring curvature becomes heavy before the displaced states have regained both the position and velocity of their moving channels. Once branch tracking is recovered, the selected states follow ordinary deterministic solutions. The calculation below explains why the light interval occurs, how quantum modes generate phase-space access, and why that access must be converted into a finite-history probability.

\section{A supergravity branch selector}
We work in reduced Planck units with chiral multiplets \(\Phi,Z\) and a nilpotent goldstino multiplet \(S^2=0\) \cite{Ferrara:2014kva,DallAgata:2014qsj}. Building on nilpotent inflationary-attractor constructions \cite{Guleryuz:2022hiv}, take
\begin{align}
 K&=K_0+fS\bar S,
 & W&=W_0+F_SS,
 \nonumber\\
 K_0&=-\frac12(\Phi-\bar\Phi)^2
      -\frac12(Z-\bar Z)^2 .
 \label{eq:KW_main}
\end{align}
The construction begins with a holomorphic equation for the desired branch geometry. Using a branch profile \({\cal V}_{\rm br}\) and an inflaton-sector function \({\cal U}_{\rm inf}\), define
\begin{align}
 {\cal B}(\Phi,Z)
 &=Z^2-{\cal V}_{\rm br}^{\,2}(\Phi),
 &
 {\cal E}
 &={\cal U}_{\rm inf}+3|W_0|^2,
 \nonumber\\
 f
 &=\frac{|F_S|^2}
 {{\cal E}(1+\gamma|{\cal B}|^2)},
 &
 \gamma&>0 .
 \label{eq:nilpotent_selector_main}
\end{align}
Already at the classical level, \({\cal B}=0\) has the two solutions \(Z_s(\Phi)=s{\cal V}_{\rm br}(\Phi)\), \(s=\pm1\), exchanged by the exact reflection \(Z\to-Z\). We prepare the ensemble on \(s=-1\), so the reflection is a symmetry of the action but not of the initial state. In particular, the positive continuation exists before any stochastic evolution is introduced; its deterministic weight from the prepared state is zero.

We use \emph{branch} for the geometric locus, \emph{channel} for its surrounding rolling region, and \emph{history} for a trajectory that ultimately reaches a recovered channel.

The inverse nilpotent metric turns this branch equation into a restoring sector. Its F-term energy separates transparently as
\begin{equation}
 \begin{aligned}
 V_S&=e^{K_0}K^{S\bar S}|D_SW|^2\\
 &=e^{K_0}{\cal E}
 +\underbrace{e^{K_0}\gamma{\cal E}|{\cal B}|^2}_{\displaystyle\Delta V_S}.
 \end{aligned}
 \label{eq:nilpotent_energy_main}
\end{equation}
The branch-dependent term has the distinctive on-branch signature
\begin{align}
 \Delta V_S\big|_{{\cal B}=0}
 &=\nabla_i\Delta V_S\big|_{{\cal B}=0}=0,
 \nonumber\\
 \left.\nabla_i\nabla_{\bar j}\Delta V_S\right|_{{\cal B}=0}
 &=e^{K_0}\gamma{\cal E}\,
 {\cal B}_{,i}\bar{\cal B}_{,\bar j}.
 \label{eq:complex_projector_main}
\end{align}
The selector contribution therefore vanishes in energy and force along either branch while retaining transverse curvature. At every regular branch point its Hermitian Hessian is an outer product of rank one: the selector curvature lies along a single complex normal direction, while tangent directions receive no curvature from \(\Delta V_S\). The branch-dependent energy is generated by the nilpotent inverse metric, rather than by an independent sgoldstino potential \cite{McDonough:2016der,Kallosh:2025dac}. Nilpotency removes the scalar of \(S\), so no propagating sgoldstino is introduced; the propagating field \(Z\) remains the branch coordinate.

For a finite selector episode, we use the shared E-model scale \(\kappa=\sqrt{2/(3\alpha)}\) and choose the explicit holomorphic profile with \( \Phi_g\equiv\phi_g/\sqrt2 \), and
\begin{equation}
 {\cal V}_{\rm br}(\Phi)
 =\frac{v_g}{\sqrt2}
 \cosh\!\left[\sqrt2\beta(\Phi-\Phi_g)\right]
 \frac{1-e^{-\sqrt2\kappa\Phi}}
      {1-e^{-\sqrt2\kappa\Phi_g}}.
 \label{eq:Vbr_main}
\end{equation}
The two factors have separate geometric roles. The \(\cosh\) factor narrows and then reopens the branch separation around the narrowing region, while the final holomorphic restoration factor merges the branches smoothly at their common scalar endpoint. This is a phenomenological tree-level profile chosen to realize a finite selector episode; no microscopic compactification origin is assumed.

The real scalar potential is therefore a consequence of this holomorphic construction, not the starting point. Choosing \({\cal U}_{\rm inf}|_{\Phi=\phi/\sqrt2}=U(\phi)\) and \(\sqrt2{\cal V}_{\rm br}(\phi/\sqrt2)=v(\phi)\), the invariant real slice \(\Phi=\phi/\sqrt2,\ Z=\tau/\sqrt2\) gives
\begin{equation}
 \begin{aligned}
 V(\phi,\tau)
 &=U(\phi)+\frac{\gamma{\cal E}(\phi)}4
 [\tau^2-v^2(\phi)]^2,\\
 U(\phi)&=M^2(1-e^{-\kappa\phi})^2,
 \qquad {\cal E}=U+3|W_0|^2,\\
 v(\phi)&=v_g\cosh[\beta(\phi-\phi_g)]
 \frac{1-e^{-\kappa\phi}}{1-e^{-\kappa\phi_g}},
 \qquad \kappa=\sqrt{\frac{2}{3\alpha}}.
 \end{aligned}
 \label{eq:V_main}
\end{equation}
Here \(U\) is the canonical E-model potential and \(M\) fixes its amplitude. The real function \(v\) is the instantaneous branch-location profile; \(v_g\), \(\beta\), and \(\phi_g\) set its scale, inverse width, and characteristic narrowing location \cite{Kallosh:2013yoa,Roest:2015qya}. Here \(\alpha\) enters through the E-model exponential scale \(\kappa\), not through the flat K\"ahler metric \(K_0\). The two real branches are \(\tau_s=s\,v(\phi)\). Their normalized selector normal, \(\hat n_{\cal B}^{(s)}=(-s\,v_{,\phi},1)/\sqrt{1+v_{,\phi}^{\,2}}\), makes the rank-one geometry explicit before the fluctuation problem is solved.

This static normal is not yet the physical entropy mass. As the inflaton rolls, \(v[\phi(N)]\) moves with e-fold time \(N=\ln a_{\rm sc}\), where \(a_{\rm sc}\) is the scale factor. A trajectory with finite response time need not sit at the instantaneous branch center. The geometric normal thereby becomes a driven dynamical direction, and the mismatch between the moving branch and the responding trajectory supplies the second ingredient of the softening.

\section{Dynamical softening without instability}
Evolving multifield backgrounds admit a standard effective-field-theory description \cite{Burgess:2012dz,Achucarro:2012sm}, but an instantaneous geometric normal does not by itself determine the physical transverse mass scale sampled by the rolling solution. That scale depends on where the trajectory sits relative to the moving center. On a fixed branch, define
\begin{equation}
 \begin{aligned}
 r&\equiv\frac{s\tau}{v}>0,
 &\ell&\equiv\frac{v'}v,\\
 A_v&\equiv\ell'+\ell^2+(3-\epsilon_H)\ell,
 &\nu&\equiv\frac{2\gamma{\cal E}v^2}{H^2},
 \end{aligned}
 \label{eq:moving_defs_main}
\end{equation}
where a prime denotes \(\dd/\dd N\) and \(\epsilon_H=-H'/H\). The ratio \(r\) locates the actual trajectory relative to the instantaneous center: \(r=1\) is perfect positional tracking. The logarithmic rate \(\ell\) and its combination \(A_v\) describe the motion and acceleration of that center, whereas \(\nu\) is its restoring stiffness in Hubble units. Substitution into the homogeneous \(\tau\) equation gives the exact moving-branch response equation
\begin{equation}
 \boxed{
 r''+(3-\epsilon_H+2\ell)r'
 +A_vr+\frac{\nu}{2}r(r^2-1)=0 .}
 \label{eq:moving_trough_main}
\end{equation}
No slow-roll or quasistatic tracking approximation is used. Equation \eqref{eq:moving_trough_main} is a driven, damped response equation: its forcing tracks the motion of the branch center, while the nonlinear final term restores the trajectory toward it. This makes the distinction between static geometry and dynamical tracking quantitative.

During the inward stage, the falling separation \(v\) lowers the \(v^2\) contribution to \(\nu\), providing the geometric source of softening. The center moves inward faster than the trajectory can respond, so the latter is initially left at \(r>1\). This inward-stage lag does not itself soften the local selector curvature: as shown below, \(r>1\) increases it relative to perfect tracking. The lag-assisted suppression arises only after the branch center turns around and begins to reopen. The trajectory responds with delay, passes to \(r<1\), and suppresses the restoring curvature while the geometric stiffness is already recovering.

This distinction is isolated by
\begin{equation}
 \varsigma\equiv\frac32(1-r^2),
 \qquad
 \frac{V_{,\tau\tau}^{\rm sel}}{H^2}
 =
 \frac{\nu}{2}(3r^2-1)
 =
 \nu(1-\varsigma).
 \label{eq:lag_curvature_main}
\end{equation}
Thus \(\varsigma\) is a signed lag correction to the branch restoring curvature. During inward motion, \(r>1\) gives \(\varsigma<0\) and hence \(1-\varsigma>1\), so finite response temporarily stiffens the selector curvature relative to perfect tracking. After the branch center turns around and reopens, the delayed trajectory passes to \(r<1\), giving \(\varsigma>0\) and suppressing the restoring fraction \(1-\varsigma\). The selector coordinate curvature vanishes locally at \(r_0=1/\sqrt3\), which we use only as its zero-curvature reference. The trajectory stays on the positive-curvature side; \(r_0\) is not a capture condition, a phase-transition criterion, or a zero of the complete entropy diagnostic.

Continuity-tracking the entropy mode connected to the selector normal \cite{Gordon:2000hv} gives the physically transparent relation
\begin{equation}
 \boxed{
 \frac{\mu_{\rm br}^2}{H^2}
 \simeq
 \nu(1-\varsigma).}
 \label{eq:lag_closure_main}
\end{equation}
The exact identity adds a remainder \(\Delta_{\rm br}/H^2\), whose relative effect at the three numerical minima is below \(5\times10^{-9}\); its derivation is given in the \cref{sec:geometry_softening}. Here \(\mu_{\rm br}^2\) is an instantaneous entropy diagnostic, not an exact eigenfrequency of the time-dependent mode system solved below.

\begin{figure}[htb!]
 \centering
 \includegraphics[width=\columnwidth]
 {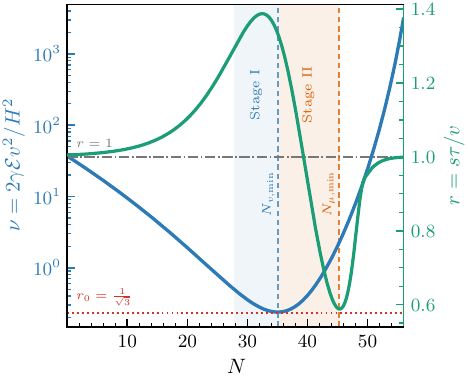}
 \caption{\textbf{Lag-assisted entropy softening} for \(\alpha=1\). Stage-I narrowing lowers the geometric stiffness; after its turnaround at \(N_{v,\min}\), Stage-II tracking lag delays the entropy minimum to \(N_{\mu,\min}\). The trajectory remains on the positive-curvature side of \(r_0=1/\sqrt3\).}
 \label{fig:mechanism}
\end{figure}

Figure~\ref{fig:mechanism} reveals the complete response. Initially, the trajectory tracks the branch center, \(r\simeq1\). As the branches narrow, their center moves inward, and the finite response leaves \(r>1\). At \(N_{v,\min}\) the separation reaches its minimum and begins to reopen, but the trajectory cannot reverse its relative motion instantaneously. Its delayed dynamical response continues under damping and the \(A_vr\) forcing term, driving \(r\) back through unity and toward \(r_0=1/\sqrt3\). By then \(\nu\) is already increasing, whereas \(1-\varsigma\) is still falling. The entropy mode is softest only after the geometric turnaround,
\begin{equation}
 N_{v,\min}<N_{\mu,\min}.
 \label{eq:separated_minima_main}
\end{equation}
This delayed ordering is the dynamical imprint of finite tracking response, rather than an independent tuning of two event times: geometric narrowing initiates the softening, whereas post-turnaround lag delays and deepens its minimum. The trajectory approaches \(r_0\) from the positive-curvature side, the tracked entropy diagnostic remains positive, and the two spectator diagnostics remain heavy. The episode is therefore heavy--light--heavy without a tachyonic instability \cite{Cormier:1999ia,Achucarro:2016fby}. At late times, the restored curvature returns \(r\) toward unity.

\section{Mode-derived quantum access}
Temporary lightness only establishes that transverse fluctuations can grow; it does not yet specify the fluctuations relevant for switching histories. That is a phase-space question. A realization may cross toward the competing channel in field space and nevertheless return because its momentum points back toward the prepared channel. A field-only \(H/(2\pi)\) kick would miss this distinction, as well as the mass dependence, mixing, phases, and field--momentum correlations inherited from the heavy--light--heavy system.

We therefore solve the complete coupled scalar modes for the four real components \(q^A=(\phi,\chi,\tau,\sigma)\) of \((\Phi,Z)\) \cite{Habib:1992ci,Grain:2017dqa,Pinol:2020cdp,Jackson:2024aoo,Christie:2026dwx}. With \({\cal Q}_k^{\cal A}=(Q_k^A,D_NQ_k^A)\), where \({\cal A}\) spans the four field and four e-fold-derivative components, the covariance injected at the coarse-graining shell \(k_\rho=\rho a_{\rm sc}H\), with fixed \(0<\rho<1\), is
\begin{equation}
 \Xi^{{\cal A}{\cal B}}(N)
 =
 (1-\epsilon_H)
 \frac{k_\rho^3}{2\pi^2}
 \sum_{\lambda=1}^{4}
 \operatorname{Re}\!\left[
 {\cal Q}_{k_\rho}^{{\cal A}\lambda}
 {\cal Q}_{k_\rho}^{{\cal B}\lambda *}
 \right].
 \label{eq:covariance_main}
\end{equation}
Only after constructing this full field--momentum covariance do we reduce to the long-wavelength phase space
\({\mathsf X}^I=(\phi,\tau,\phi',\tau')\):
\begin{equation}
 \begin{aligned}
 \dd{\mathsf X}^I
 &=B^I({\mathsf X})\,\dd N
 +\Sigma^I{}_a(N)\,\dd W_N^a,\\
 \Sigma\Sigma^{\mathsf T}&=\Xi_{\rm ret}.
 \end{aligned}
 \label{eq:SDE_main}
\end{equation}
Here \(B^I\) is the nonlinear classical drift, \(\dd W_N^a\) are independent Wiener increments, and \(\Xi_{\rm ret}\) is the marginal covariance retained on this reduced phase space; \(\Sigma\) is its mode-derived noise factor. Thus the stochastic field, momentum, and field--momentum cross-correlations are inherited from the same coupled scalar perturbation problem that contains the temporarily soft entropy sector. The soft mode and the source that probes the switching region are therefore not introduced as independent phenomenological approximations.

During the light interval, the accumulated transverse width becomes comparable to the nonlinear switching distance between outgoing channels. The covariance-whitened local accessibility defined in the \cref{sec:stochastic_selection} reaches
\begin{equation}
 Q_{\rm acc}^{\perp,\max}=0.91\text{--}1.11,
 \label{eq:accessibility_main}
\end{equation}
across the three realizations. Values of order unity mean that the local deterministic switching distance has become comparable to the accumulated transverse covariance scale. They do not give a branch probability. A realization can enter the competing side locally and later return, especially when its velocity is directed back toward the original channel. Local accessibility must therefore be followed by a genuinely finite-history question: which recovered heavy channel is reached first?

\section{Finite-history selection}
Local access is not yet history selection. A trajectory may cross \(\tau=0\), or even enter the competing side of the local switching region, and subsequently return to the prepared channel. We count a history as selected only when it reaches an outgoing heavy channel and has recovered both its configuration and velocity tracking. The two possible outcomes are therefore recovered positive and negative classical continuations, rather than the two signs of an instantaneous field displacement.

For a finite horizon \(N_f\), \(p_s^{\rm sel}\) denotes the conditional probability that a history resolves on continuation \(s=\pm1\), among histories that reach either recovered continuation before leaving the controlled domain. This is a finite-horizon first-passage probability \cite{Vennin:2015hra,Assadullahi:2016gkk}, not a global cosmological measure. Related finite-horizon survival problems use an absorbing control boundary \cite{Guleryuz:2026rdz}; here the competing endpoints are recovered inflationary channels. In the stochastic ensemble, its estimator has the transparent form
\begin{equation}
 \boxed{
 \widehat p_s^{\rm sel}
 =
 \frac{n_s}{n_++n_-},
 }
 \label{eq:committor_main}
\end{equation}
where \(n_s\) counts histories first recovered on continuation \(s\). Histories lost from the controlled domain or unresolved at \(N_f\) do not enter the conditional denominator. Selection therefore requires a completed dynamical transition into a recovered classical channel, not merely local covariance-scale access to it.

The origin of this weight can now be tested by a controlled counterfactual. We deform only the mode-derived source, \(\Sigma\to\varepsilon_q\Sigma\), \(0\le\varepsilon_q\le1\), while holding fixed the classical action, prepared initial state, nonlinear drift \(B^I\), recovery criterion, controlled domain, selection rule, and finite horizon. For the central \(\alpha=1\) realization,
\begin{equation}
 p_+^{\rm sel}(0)=0,
 \qquad
 \widehat p_+^{\rm sel}(1)=0.3705 .
 \label{eq:sharp_result_main}
\end{equation}
The zero-source equality is the exact deterministic initial-value result for the same prepared state, which remains on the negative branch; it is not an extrapolation from finite noise. The physical-source estimate uses \(2\times10^4\) histories and has a \(95\%\) Wilson interval \(0.3638<p_+^{\rm sel}<0.3772\). Across the three CMB-normalized realizations, \(\widehat p_+^{\rm sel}=0.3343\)--\(0.3957\).
\begin{figure}[htb!]
 \centering
 \includegraphics[width=\columnwidth]{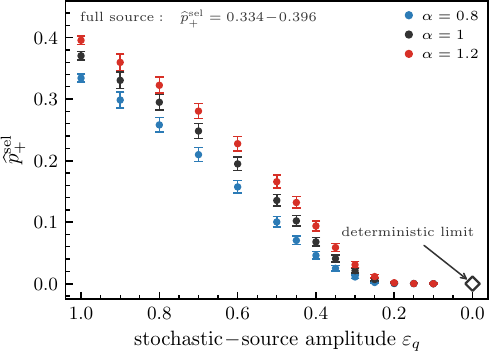}
 \caption{\textbf{Quantum origin of the competing-history weight.} Estimated conditional positive-history weight \(\widehat p_+^{\rm sel}\) as the mode-derived stochastic source is suppressed, \(\Sigma\to\varepsilon_q\Sigma\). All three realizations approach the exact deterministic zero as \(\varepsilon_q\to0\).}
 \label{fig:noise_scan}
\end{figure}

Figure~\ref{fig:noise_scan} supplies the counterfactual evidence for quantum accessibility. The competing weight decreases smoothly across the scan as the source is suppressed and terminates at the exact deterministic zero. Nothing in this deformation creates or removes the positive classical continuation. Rather, the continuation is classical in its existence and subsequent motion, but quantum in its accessibility from the prepared state within the stated stochastic closure.

\section{Classical recovery}
Selection is not yet the end of the argument: a history populated by quantum fluctuations need not require continued stochastic support. After the light interval, the branch mode re-heavies. The reference background subsequently satisfies a horizon-persistent heavy-sector recovery gate: \(N_{\rm heavy}\) denotes the first post-light gate that remains satisfied through the audit horizon \(N_f=54\), while \(N_{\rm track}\) marks the later recovery of configuration and velocity tracking. Individual stochastic histories satisfy the same recovery criterion at their own pathwise recovery times. Figure~\ref{fig:dynamics} separates the two reference markers, \(N_{\rm heavy}<N_{\rm track}\), with \(N_{\rm track}-N_{\rm heavy}=0.41\)--\(0.46\) across the three realizations. This ordering has a direct dynamical interpretation: the heavy sector can recover before the reference trajectory has had time to relax onto the moving channel.

A direct recovery test starts with representative captured stochastic states from their exact phase-space coordinates and then sets \(\Sigma=0\). No state is projected onto a branch, no velocity is reset, and no trajectory is matched by hand. Nevertheless, the positive and negative samples approach their corresponding deterministic outgoing solutions in Fig.~\ref{fig:dynamics}(a). Quantum fluctuations are therefore needed to select the history from the prepared state, but not to sustain the selected continuation: once captured, its subsequent motion is supported by the ordinary classical equations.

\section{Discussion}
Taken together, the dynamics form a single causal chain. The branch geometry narrows; finite response turns its motion into lag-assisted softening; the coupled quantum modes supply a phase-space source; first passage converts temporary access into a weight over recovered continuations; and recovery of the heavy regime followed by branch tracking returns both outcomes to deterministic motion. The histories in Fig.~\ref{fig:dynamics} exhibit the full evolution, the lag dynamics in Fig.~\ref{fig:mechanism} explain why the stable mode becomes light, and the source deformation in Fig.~\ref{fig:noise_scan} isolates the quantum origin of the competing weight. Classical uniqueness is never violated. The quantum episode changes the probability of reaching solutions that were already present in the classical theory.

History weights and local perturbative observables consequently need not carry the same imprint. The former are finite-history first-passage probabilities built from field, momentum, and their cross-correlations. For the tested modes exiting before the selector episode, no selector-induced curvature transfer is resolved despite an order-\(10^{-1}\) redistribution between outgoing histories. This separation is natural: a substantial change in which classical endpoint is reached need not appear as a comparably large branch-even local two-point signal. Later branch-dependent physics could remain sensitive to the unequal weights.

The construction has a deliberately limited scope. It is a tree-level bosonic scalar EFT of constrained supergravity; the microscopic origin of the nilpotent multiplet, fermionic and gravitino dynamics, and radiative stability remain open. The noise covariance is computed on the rolling reference background rather than from a separate ultraviolet mode problem along each nonlinear realization, defining a background-derived Markov closure. A central sharp-shell scan retains the competing-history population but shifts its numerical weight, so coarse-graining invariance is not claimed. Decoherence, Born-rule interpretation, cosmological measure theory, and non-Markovian or open-system corrections lie outside the calculation \cite{Christie:2026dwx,Cruces:2026yvs,Li:2026lwl}. The nearby \(\alpha\) realizations likewise establish persistence within a coordinated CMB-normalized window, not a universal scaling law.

What matters beyond the particular selector profile is the dynamical sequence it realizes: pre-existing classical continuations, finite quantum access between them, and subsequent recovery of deterministic evolution. Inflation provides a cosmological laboratory in which this sequence can be realized and followed explicitly, but none of these ingredients is intrinsically inflationary. It is therefore natural to ask whether the same mechanism can arise more generally in cosmological or quantum field-theoretic systems with transiently accessible classical alternatives. A finite fluctuation episode can end while its consequence survives as a redistribution among subsequently classical futures.

\appendix

\section{Derivations and Numerical Certification}\label{sec:Appendix}
This appendix gives the derivations and numerical certification underlying the paper. We derive the nilpotent-supergravity selector and exact moving-branch dynamics, specify the background and coupled-mode evolution, construct the mode-derived stochastic phase-space covariance and finite-history selection problem, and summarize the numerical and stochastic-closure tests. We finally test curvature two-point transfer for modes exiting before the selector episode. Reduced Planck units \(M_{\rm Pl}=1\) are used throughout, \(N=\ln a_{\rm sc}\), and a prime denotes \(\dd/\dd N\). We use ``branch'' for the geometric locus \({\cal B}=0\), ``channel'' for its associated rolling region, and ``history'' or ``continuation'' for a dynamical solution that ultimately occupies one of the recovered channels.

\subsection{Nilpotent-SUGRA construction}
\label{sec:kahler}
The representative used in the paper is
\begin{align}
 K&=K_0+fS\bar S,
 \qquad
 W=W_0+F_SS,
 \qquad
 S^2=0,
 \label{eq:KW_supp}\\
 K_0&=
 -\frac12(\Phi-\bar\Phi)^2
 -\frac12(Z-\bar Z)^2 .
 \nonumber
\end{align}
The nilpotent metric factor is chosen as
\begin{equation}
 f=
 \frac{|F_S|^2}
 {{\cal E}(1+\gamma|{\cal B}|^2)},
 \qquad
 \gamma>0,
 \label{eq:f_supp}
\end{equation}
with
\begin{align}
 {\cal E}&={\cal U}_{\rm inf}+3|W_0|^2,
 &
 {\cal B}&=Z^2-{\cal V}_{\rm br}^{\,2}(\Phi),
 \label{eq:EB_supp}\\
 {\cal U}_{\rm inf}
 &=M^2\left|1-e^{-\sqrt2\kappa\Phi}\right|^2,
 &
 \kappa&=\sqrt{\frac{2}{3\alpha}},
 \nonumber
\end{align}
and
\begin{equation}
 {\cal V}_{\rm br}(\Phi)
 =
 \frac{v_g}{\sqrt2}
 \cosh\!\left[\sqrt2\beta(\Phi-\Phi_g)\right]
 \frac{1-e^{-\sqrt2\kappa\Phi}}
      {1-e^{-\sqrt2\kappa\Phi_g}} .
 \label{eq:profiles_holo_supp}
\end{equation}
The unconstrained \((\Phi,Z)\) sector is flat. Thus \(\alpha\) enters through the canonical E-model exponential scale \(\kappa\), not through a curved \(\alpha\)-attractor K\"ahler manifold; the construction is a supergravity realization of an E-model-shaped scalar potential rather than an identification with the conventional curved \(\alpha\)-attractor geometry.

At \(S=0\),
\begin{equation}
 \begin{aligned}
 K^{S\bar S}|D_SW|^2
 &={\cal E}(1+\gamma|{\cal B}|^2),\\
 D_iW&=K_{0,i}W_0\qquad(i=\Phi,Z).
 \end{aligned}
 \label{eq:nilpotent_chain_supp}
\end{equation}
and hence
\begin{equation}
 V_F
 =
 e^{K_0}
 \left[
 {\cal E}(1+\gamma|{\cal B}|^2)
 +|W_0|^2G^{i\bar j}K_{0,i}K_{0,\bar j}
 -3|W_0|^2
 \right].
 \label{eq:Fterm_supp}
\end{equation}
We decompose the two unconstrained complex scalars into four real fields as
\begin{equation}
 \Phi=\frac{1}{\sqrt2}(\phi+i\chi),
 \qquad
 Z=\frac{1}{\sqrt2}(\tau+i\sigma).
 \label{eq:real_decomposition_supp}
\end{equation}
The homogeneous reference background lies on the invariant real slice
\begin{equation}
 \chi=\sigma=0,
 \qquad
 \Phi=\frac{\phi}{\sqrt2},
 \qquad
 Z=\frac{\tau}{\sqrt2}.
 \label{eq:real_slice_supp}
\end{equation}
The orthogonal fields \(\chi\) and \(\sigma\) are not discarded: their fluctuations are retained in the complete four-real-field perturbation system and enter the entropy-spectrum and stochastic-covariance analyses below. On the background slice, \(K_0=K_{0,i}=0\), and
\begin{equation}
 \begin{aligned}
 {\cal B}
 &=\frac12[\tau^2-v^2(\phi)],\\
 v(\phi)
 &=v_g\cosh[\beta(\phi-\phi_g)]
 \frac{1-e^{-\kappa\phi}}
      {1-e^{-\kappa\phi_g}} .
 \end{aligned}
 \label{eq:v_profile_supp}
\end{equation}
Defining
\begin{equation}
 U(\phi)=M^2(1-e^{-\kappa\phi})^2,
 \qquad
 {\cal E}(\phi)=U(\phi)+3|W_0|^2,
 \label{eq:UE_real_supp}
\end{equation}
Eq.~\eqref{eq:Fterm_supp} reduces on this slice to
\begin{equation}
 \boxed{
 V(\phi,\tau)
 =
 U(\phi)
 +
 \frac{\gamma{\cal E}(\phi)}4
 [\tau^2-v^2(\phi)]^2 .}
 \label{eq:V_rederived_supp}
\end{equation}
This reproduces Eq.~\eqref{eq:V_main} of the paper. The two background branches are
\begin{equation}
 \tau_s(\phi)=s\,v(\phi),
 \qquad
 s=\pm1.
 \label{eq:real_branches_supp}
\end{equation}

The branch-dependent contribution
\begin{equation}
 \Delta V_S=e^{K_0}\gamma{\cal E}|{\cal B}|^2
\end{equation}
and its first derivatives vanish on \({\cal B}=0\), while
\begin{equation}
 \left.
 \nabla_i\nabla_{\bar j}\Delta V_S
 \right|_{{\cal B}=0}
 =
 e^{K_0}\gamma{\cal E}\,
 {\cal B}_{,i}\bar{\cal B}_{,\bar j}.
 \label{eq:complex_projector_supp}
\end{equation}
At regular branch points \({\cal B}_{,i}\neq0\), this Hermitian \((\Phi,Z)\) block has rank one and selects one complex normal direction. Its restriction to the invariant real plane gives the rank-one real selector Hessian derived below. This does not imply a unique real normal in the full four-real-field fluctuation space; the physical entropy hierarchy follows from the complete coupled scalar system.

Finally,
\begin{equation}
 U(\phi)\to0,
 \qquad
 v(\phi)\to0
 \qquad
 (\phi\to0),
 \label{eq:restoration_supp}
\end{equation}
so \(\tau_\pm=\pm v(\phi)\) merge smoothly at \((\phi,\tau)=(0,0)\), where \(V(0,0)=0\). The selector therefore leaves no permanent scalar barrier at the common endpoint.

\subsection{Moving-branch dynamics and entropy softening}
\label{sec:geometry_softening}
Define
\begin{equation}
 X(\phi,\tau)=\tau^2-v^2(\phi),
 \qquad
 V_{\rm sel}=\frac{\gamma{\cal E}}4X^2.
\end{equation}
On the instantaneous branch locus \(X=0\),
\begin{equation}
 \left.
 \partial_i\partial_jV_{\rm sel}
 \right|_{X=0}
 =
 \frac{\gamma{\cal E}}2
 X_{,i}X_{,j},
 \label{eq:rank_one_hessian_supp}
\end{equation}
so the selector Hessian on the real \((\phi,\tau)\) plane has rank one at regular branch points. Away from the branch center, where the rolling trajectory generally has \(X\neq0\), the full selector Hessian need not remain rank one.

For \(\tau_s=s\,v(\phi)\), \(s=\pm1\),
\begin{equation}
 \hat t_s
 =
 \frac{(1,s\,v_{,\phi})}
 {\sqrt{1+v_{,\phi}^2}},
 \qquad
 \hat n_{\cal B}^{(s)}
 =
 \frac{(-s\,v_{,\phi},1)}
 {\sqrt{1+v_{,\phi}^2}},
 \label{eq:tangent_normal_supp}
\end{equation}
form an orthonormal tangent--normal pair, with
\begin{equation}
 H_{\rm sel}\hat t_s=0,
 \qquad
 H_{\rm sel}\hat n_{\cal B}^{(s)}
 =
 m_{\perp,\rm sel}^2\hat n_{\cal B}^{(s)},
\end{equation}
and
\begin{equation}
 m_{\perp,\rm sel}^2
 =
 2\gamma{\cal E}v^2(1+v_{,\phi}^2).
 \label{eq:normal_mass_supp}
\end{equation}
The corresponding \(\tau\)-coordinate curvature is
\begin{equation}
 m_{\tau,\rm tr}^2
 =
 V_{,\tau\tau}^{\rm sel}\big|_{\tau=sv}
 =
 2\gamma{\cal E}v^2.
 \label{eq:mtr_supp}
\end{equation}
Thus \(m_{\perp,\rm sel}^2\) is the nonzero selector-Hessian eigenvalue on the branch, while \(m_{\tau,\rm tr}^2\) is the coordinate curvature controlling the moving-branch dynamics.

On an interval with \(v\neq0\) and fixed sign of \(\tau\), define
\begin{equation}
 r(N)=\frac{s\tau(N)}{v[\phi(N)]}>0,
 \qquad
 \ell=\frac{v'}v .
 \label{eq:r_signed_supp}
\end{equation}
Because the base potential is \(\tau\)-independent,
\begin{equation}
 \tau''+(3-\epsilon_H)\tau'
 +\frac{\gamma{\cal E}}{H^2}\tau(\tau^2-v^2)=0,
 \label{eq:tau_background_supp}
\end{equation}
and substitution of \(\tau=svr\) gives
\begin{equation}
 \boxed{
 r''+(3-\epsilon_H+2\ell)r'
 +A_vr+\frac{\nu}{2}r(r^2-1)=0,}
 \label{eq:moving_trough_supp}
\end{equation}
with
\begin{equation}
 A_v=\ell'+\ell^2+(3-\epsilon_H)\ell,
 \qquad
 \nu=\frac{2\gamma{\cal E}v^2}{H^2}.
 \label{eq:moving_coeffs_supp}
\end{equation}
No slow-roll or quasistatic tracking approximation is used. The term \(A_vr\) encodes the kinematic forcing required to follow the moving branch center, while the last term is the nonlinear selector restoring force.

Introduce
\begin{equation}
 \varsigma
 =
 \frac32(1-r^2),
 \label{eq:softness_supp}
\end{equation}
for which the selector coordinate curvature along the rolling trajectory is exactly
\begin{equation}
 \boxed{
 \frac{V_{,\tau\tau}^{\rm sel}}{H^2}
 =
 \frac{\nu}{2}(3r^2-1)
 =
 \nu(1-\varsigma).}
 \label{eq:lag_mass_supp}
\end{equation}
Perfect tracking gives \(r=1\) and \(\varsigma=0\). In the present background, finite response appears with opposite effects on the two sides of the geometric turnaround. While the branch center moves inward, \(r>1\) and hence \(\varsigma<0\), so the lag correction stiffens the local selector curvature relative to perfect tracking. During the delayed response to reopening, \(r<1\) and hence \(\varsigma>0\), so the lag correction instead suppresses that curvature. The local coordinate curvature vanishes at
\begin{equation}
 r_0=\frac1{\sqrt3},
 \qquad
 \varsigma=1.
\end{equation}
This point is used only as a local positive-curvature-side reference: it is neither a capture condition nor a phase-transition criterion, and its coordinate-curvature zero does not imply a zero of the complete entropy diagnostic. The certified trajectories approach \(r_0\) from above without crossing it.

To connect Eq.~\eqref{eq:lag_mass_supp} to the physical entropy hierarchy, let
\begin{equation}
 T^A=\frac{q^{A\prime}}{\sqrt{q^{B\prime}q'_B}}
\end{equation}
be the adiabatic tangent, \(N_a^{\,A}\) an orthonormal entropy frame, and
\begin{equation}
 \Omega_a=N_{aA}D_NT^A
\end{equation}
the entropy-space turn vector. The projected Mukhanov--Sasaki operator and instantaneous entropy diagnostic are
\begin{align}
 {\cal M}^{\perp}_{ab}
 &=
 N_{aA}{\cal M}^{A}{}_{B}N_b^{\,B},
 \nonumber\\
 \mu^2_{ab}
 &=
 {\cal M}^{\perp}_{ab}
 +
 3H^2\Omega_a\Omega_b .
 \label{eq:entropy_definition_supp}
\end{align}
Its eigenvalues diagnose the instantaneous normal hierarchy; they are not identified with exact eigenfrequencies of the full time-dependent coupled system.

For the present flat canonical field space, the explicit field-space curvature term in \({\cal M}^{A}{}_{B}\) vanishes. Double projection also removes its velocity-dependent gravitational term because \(N_{aA}q^{A\prime}=0\). If \(\hat e_{\rm br}\) is the continuity-tracked entropy eigenvector connected to the selector-normal sector,
\begin{equation}
 \frac{\mu_{\rm br}^2}{H^2}
 =
 \frac{
 \hat e_{\rm br}^{\,T}\nabla\nabla V_{\rm sel}\hat e_{\rm br}
 }{H^2}
 +
 \frac{
 \hat e_{\rm br}^{\,T}\nabla\nabla V_{\rm base}\hat e_{\rm br}
 }{H^2}
 +
 3(\hat e_{\rm br}\!\cdot\!\bm{\Omega})^2 .
 \label{eq:exact_entropy_decomp_supp}
\end{equation}
Define
\begin{align}
 \frac{\Delta_{\rm br}}{H^2}
 &\equiv
 \frac{
 \hat e_{\rm br}^{\,T}\nabla\nabla V_{\rm sel}\hat e_{\rm br}
 -
 V_{,\tau\tau}^{\rm sel}
 }{H^2}
 \nonumber\\
 &\quad+
 \frac{
 \hat e_{\rm br}^{\,T}\nabla\nabla V_{\rm base}\hat e_{\rm br}
 }{H^2}
 +
 3(\hat e_{\rm br}\!\cdot\!\bm{\Omega})^2 ,
 \label{eq:delta_branch_supp}
\end{align}
so that
\begin{equation}
 \boxed{
 \frac{\mu_{\rm br}^2}{H^2}
 =
 \nu(1-\varsigma)
 +
 \frac{\Delta_{\rm br}}{H^2}.}
 \label{eq:exact_lag_decomp_supp}
\end{equation}
This identity concerns the adopted instantaneous entropy diagnostic, not an exact normal-mode frequency of the time-dependent perturbation system.

At the three numerical minima, omitting \(\Delta_{\rm br}\) changes the complete branch-entropy diagnostic by less than \(5\times10^{-9}\) relatively. Including the full remainder, Eq.~\eqref{eq:exact_lag_decomp_supp} closes with maximum absolute residual \(4.26\times10^{-14}\) across the three backgrounds. Thus the minimum is quantitatively exhausted, to numerical precision, by the selector-sector factor \(\nu(1-\varsigma)\). Geometric narrowing first reduces \(\nu\); after the geometric turnaround, delayed tracking suppresses \(1-\varsigma\) and shifts the minimum to later times, while the complete branch diagnostic remains nontachyonic.

\begin{table}[tbp]
\centering
\caption{Lag-induced softening of the continuity-tracked branch entropy diagnostic; it remains positive in all three realizations.}
\label{tab:spinodal_supp}
\begin{tabular}{@{}lccc@{}}
\toprule
 & \(\alpha=0.8\) & \(\alpha=1.0\) & \(\alpha=1.2\)\\
\midrule
\(N_{v,\min}\)
& \(35.0381\) & \(35.0477\) & \(35.0572\)\\
\(N_{\mu,\min}\)
& \(45.1419\) & \(45.2407\) & \(45.3327\)\\
\(N_{\mu,\min}-N_{v,\min}\)
& \(10.1038\) & \(10.1930\) & \(10.2755\)\\
\(\min(\mu_{\rm br}^2/H^2)\)
& \(0.0329\) & \(0.0422\) & \(0.0504\)\\
\(r(N_{\mu,\min})\)
& \(0.5856\) & \(0.5882\) & \(0.5906\)\\
\(\varsigma(N_{\mu,\min})\)
& \(0.9855\) & \(0.9810\) & \(0.9767\)\\
Selector-curvature crossing
& no & no & no\\
\bottomrule
\end{tabular}
\end{table}

We define the diagnostic light interval by
\begin{equation}
 \frac{\mu_{\rm br}^2}{H^2}<1.
 \label{eq:light_definition_supp}
\end{equation}
Equation~\eqref{eq:lag_mass_supp} reproduces its numerical boundaries to the displayed precision in all three realizations. This serves as a closure test that the temporary branch-mode softening is controlled by the lag-suppressed selector sector, not as an independent precision observable.

\subsection{Parameters, background, and coupled-mode evolution}
\label{sec:background_modes}
The parameters common to the three primary realizations are
\begin{equation}
 \begin{aligned}
 \gamma&=5.1681\times10^{10},
 &v_g&=8.7672\times10^{-7},\\
 W_0&=10^{-10},
 &F_S&=1,
 \qquad N_g=35 .
 \end{aligned}
 \label{eq:common_parameters_supp}
\end{equation}
The normalization \(F_S\) cancels from \(K^{S\bar S}|D_SW|^2\) and does not independently enter the real-slice potential. Although \(\gamma\) is large in Planck units,
\begin{equation}
 \gamma v_g^2\simeq3.97\times10^{-2},
\end{equation}
while the dynamical selector scales are more directly measured by \(\nu=2\gamma{\cal E}v^2/H^2\) and \(\mu_{\rm br}^2/H^2\).

The deformation is fixed by
\begin{equation}
 \begin{aligned}
 \beta(\alpha)&=\frac{3.26}{\sqrt{\alpha}},
 &\epsilon_V(\phi_{\rm end})&=1,\\
 N_{\rm SR}(\phi_g)&=25,
 &N_{\rm SR}(\phi_{\rm start})&=70 .
 \end{aligned}
 \label{eq:campaign_construction_supp}
\end{equation}
where \(N_{\rm SR}(\phi)=\int_{\phi_{\rm end}}^\phi U/U_{,\phi}\,\dd\phi\).  These equations, together with the common parameters above, generate the \(\beta\), \(\phi_g\), and \(\phi_{\rm start}\) entries in Table~\ref{tab:campaign_parameters_supp}; no parameter is fitted separately to the stochastic outcome.

\begin{table}[tbp]
\centering
\caption{Parameters of the three primary E-model realizations. The amplitude \(M\) is fixed using the complete coupled-mode scalar power at the common internal normalization mode.}
\label{tab:campaign_parameters_supp}
\begin{tabular}{@{}lccc@{}}
\toprule
 & \(\alpha=0.8\) & \(\alpha=1.0\) & \(\alpha=1.2\)\\
\midrule
\(M\)
& \(9.3791\times10^{-6}\)
& \(1.0414\times10^{-5}\)
& \(1.1337\times10^{-5}\)\\
\(\beta\)
& \(3.6448\)
& \(3.2600\)
& \(2.9760\)\\
\(\phi_g\)
& \(4.2171\)
& \(4.4669\)
& \(4.6745\)\\
\(\phi_{\rm start}\)
& \(5.2710\)
& \(5.6318\)
& \(5.9372\)\\
\bottomrule
\end{tabular}
\end{table}

For each realization, the e-fold coordinate is shifted so that
\begin{equation}
 \phi(N_g)=\phi_g,
 \qquad
 N_g=35.
 \label{eq:Ng_alignment_supp}
\end{equation}
Because the E-model restoration factor also enters \(v(\phi)\), the point \(\phi_g\), defined by \(v(\phi_g)=v_g\), is slightly displaced from the exact minimum of the branch separation. All numerical calculations use the complete profile.

Writing
\begin{equation}
 q^A=(\phi,\chi,\tau,\sigma),
\end{equation}
the homogeneous equations are
\begin{align}
 q^{A\prime\prime}
 +(3-\epsilon_H)q^{A\prime}
 +\frac{V^{,A}}{H^2}
 &=0,
 \nonumber\\
 \epsilon_H
 &=
 \frac12q_A'q^{A\prime},
 &
 H^2
 &=
 \frac{V}{3-\epsilon_H}.
 \label{eq:background_supp}
\end{align}
The reflection-symmetric slice \(\chi=\sigma=0\) is dynamically invariant. The reference history is initialized on the negative branch:
\begin{align}
 \phi_0&=\phi_{\rm start},
 &
 \tau_0&=-v(\phi_0),
 \nonumber\\
 \chi_0&=\sigma_0=0,
 \nonumber\\
 u_{\phi,0}
 &=
 -\frac{V_{,\phi}(\phi_0,\tau_0)}
 {V(\phi_0,\tau_0)},
 &
 u_{\tau,0}
 &=
 -v_{,\phi}(\phi_0)u_{\phi,0},
 \nonumber\\
 u_{\chi,0}&=u_{\sigma,0}=0 .
 \label{eq:background_initial_supp}
\end{align}
These relations specify only the prepared initial state; subsequent evolution uses Eq.~\eqref{eq:background_supp} without imposing slow roll. The prepared ensemble is therefore not reflection symmetric despite the exact symmetry between the two branches.

For each \(\alpha\), \(M\) is fixed using the complete coupled four-field perturbation system. The common internal normalization mode is labelled by its horizon-crossing time \(N_k=5\), defined by
\begin{equation}
 k_5=a_{\rm sc}(5)H(5).
\end{equation}
Its curvature power is evaluated at the common reporting time \(N_{\rm pre}=33\), and we impose
\begin{equation}
 {\cal P}_{\cal R}(k_5,N_{\rm pre}=33)=A_s^{\rm obs},
 \qquad
 A_s^{\rm obs}=2.10\times10^{-9}.
 \label{eq:cmb_norm_supp}
\end{equation}
Starting from the slow-roll amplitude estimate, the full background and mode system is recomputed under the iteration
\begin{equation}
 M_{j+1}=M_j
 \sqrt{\frac{A_s^{\rm obs}}
 {{\cal P}_{\cal R}(k_5,33;M_j)}} .
 \label{eq:M_iteration_supp}
\end{equation}
until the reported relative normalization tolerance is reached. The resulting amplitudes agree with \(A_s^{\rm obs}\) to better than \(1.5\times10^{-4}\) relatively. This fixes the inflationary amplitude to the observed scalar-amplitude scale, but \(N_k=5\) is only a common pre-selector internal reference mode and is not identified with a present-day pivot wavenumber without specifying reheating.

Entropy eigenvectors are evaluated on a uniform 220-point grid over \(12\le N\le52\). Between adjacent slices, the permutation maximizing the sum of absolute canonical overlaps is chosen, and signs are fixed continuously. A final one-to-one assignment to the branch, \(\chi\), and \(\sigma\) directions maximizes their mean squared overlaps over the grid. Cubic splines of these continuity-tracked eigenvalues locate minima and threshold crossings. The spectator minima are also checked on a 360-point extension from \(N=52\) to the numerical end of inflation, \(\epsilon_H=1\). Thus the labels are not reset by instantaneous eigenvalue ordering during the temporary softening.

\begin{table}[tbp]
\centering
\caption{Continuity-tracked entropy hierarchy. Spectator minima are taken
over the complete inflationary interval \(\epsilon_H<1\).}
\label{tab:selector_summary_supp}
\begin{tabular}{@{}lccc@{}}
\toprule
 & \(\alpha=0.8\) & \(\alpha=1.0\) & \(\alpha=1.2\)\\
\midrule
\(\min(\mu_{\rm br}^2/H^2)\)
& \(0.0329\) & \(0.0422\) & \(0.0504\)\\
\(\min(\mu_\chi^2/H^2)\)
& \(6.059\) & \(6.059\) & \(6.058\)\\
\(\min(\mu_\sigma^2/H^2)\)
& \(6.318\) & \(6.317\) & \(6.315\)\\
\(\Delta N_{\rm light}\)
& \(19.1027\) & \(19.4157\) & \(19.7084\)\\
\bottomrule
\end{tabular}
\end{table}

Exactly one continuity-tracked entropy mode enters \(\mu^2/H^2<1\) for a finite interval while remaining positive; the two orthogonal spectator diagnostics remain heavy throughout inflation. The stochastic access studied below therefore originates from a single temporarily light, nontachyonic branch sector, not from a waterfall instability or loss of spectator stabilization.

The stochastic source is derived from the complete coupled mode system rather than assigned through a massless \(H/(2\pi)\) prescription. The perturbations obey
\begin{equation}
 D_N^2Q_k^A+(3-\epsilon_H)D_NQ_k^A
 +\left[
 \frac{k^2}{a_{\rm sc}^2H^2}\delta^A{}_B
 +\frac{{\cal M}^{A}{}_{B}}{H^2}
 \right]Q_k^B=0 ,
 \label{eq:MS_supp}
\end{equation}
with
\begin{equation}
 \begin{aligned}
 {\cal M}^{A}{}_{B}
 &=G^{AC}\nabla_B\nabla_CV
 -{\cal R}^{A}{}_{CDB}\dot q^C\dot q^D\\
 &\quad-\frac{1}{a_{\rm sc}^3}
 D_t\!\left(
 \frac{a_{\rm sc}^3}{H}\dot q^A\dot q_B
 \right).
 \end{aligned}
 \label{eq:mass_operator_supp}
\end{equation}
Here a dot denotes \(\dd/\dd t\); the explicit field-space curvature term vanishes for the present flat canonical metric.

For each \(k\), four independent positive-frequency solutions are initialized at
\begin{equation}
 x_i\equiv\frac{k}{a_{\rm sc}H}=80
\end{equation}
in an orthonormal field basis \(e_\lambda^A\):
\begin{align}
 Q_{k\lambda}^A
 &=
 \frac{e_\lambda^A}{a_{\rm sc}\sqrt{2k}},
 \nonumber\\
 P_{k\lambda}^A
 &=
 -(1+ix_i)Q_{k\lambda}^A,
 \qquad
 P_k^A\equiv D_NQ_k^A .
 \label{eq:mode_initial_supp}
\end{align}
The factor \(1+ix_i\) includes the e-fold derivative of both the positive-frequency phase and the \(a_{\rm sc}^{-1}\) prefactor. Canonical normalization is monitored through
\begin{equation}
 a_{\rm sc}^3H
 \left(
 QP^\dagger-Q^*P^{\mathsf T}
 \right)
 =
 i\mathbf1 ,
 \label{eq:wronskian_supp}
\end{equation}
whose maximum relative Frobenius residual is \(1.023\times10^{-8}\) across the three primary realizations.

\subsection{Mode-derived stochastic reduction and finite-history selection}
\label{sec:stochastic_selection}
The reduced nonlinear phase-space state is
\begin{equation}
 {\mathsf X}=(\phi,\tau,u_\phi,u_\tau),
 \qquad
 u_\phi=\phi',
 \qquad
 u_\tau=\tau',
\end{equation}
with drift
\begin{align}
 \epsilon_H
 &=
 \frac12(u_\phi^2+u_\tau^2),
 &
 H^2
 &=
 \frac{V}{3-\epsilon_H},
 \nonumber\\
 B^\phi&=u_\phi,
 &
 B^\tau&=u_\tau,
 \nonumber\\
 B^{u_\phi}
 &=
 -(3-\epsilon_H)u_\phi-\frac{V_{,\phi}}{H^2},
 &
 B^{u_\tau}
 &=
 -(3-\epsilon_H)u_\tau-\frac{V_{,\tau}}{H^2}.
 \label{eq:drift_supp}
\end{align}

At each shell crossing, the complete eight-dimensional field--momentum covariance is assembled before reduction. For
\begin{equation}
 {\cal Q}_k^{\cal A}=(Q_k^A,P_k^A),
 \qquad
 P_k^A=D_NQ_k^A,
\end{equation}
we use
\begin{equation}
 \begin{aligned}
 \Xi^{{\cal A}{\cal B}}(N)
 &=(1-\epsilon_H)\frac{k_\rho^3}{2\pi^2}
 \sum_{\lambda=1}^{4}
 \operatorname{Re}\!\left[
 {\cal Q}_{k_\rho}^{{\cal A}\lambda}
 {\cal Q}_{k_\rho}^{{\cal B}\lambda *}
 \right],\\
 k_\rho&=\rho a_{\rm sc}H,
 \qquad \rho=0.10 .
 \end{aligned}
 \label{eq:full_covariance_supp}
\end{equation}
The factor \(1-\epsilon_H\) follows from \(\dd\ln k_\rho/\dd N=1-\epsilon_H\). This covariance retains the field, momentum, and field--momentum correlations of the coupled quantum system. The kernel is computed at the 220 shell times
\begin{equation}
 N_j=20+\frac{30j}{219},
 \qquad j=0,\ldots,219,
 \label{eq:shell_grid_supp}
\end{equation}
with \(k_j=\rho a_{\rm sc}(N_j)H(N_j)\); every mode is initialized at \(k_j/(a_{\rm sc}H)=80\). Each independent component of the symmetric covariance is interpolated with a not-a-knot cubic spline. At every evaluation the matrix is symmetrized, its floating-point negative eigenvalues are clipped to zero, and \(\Sigma\) is the symmetric spectral square root. For \(50<N\le54\) the endpoint value \(\Sigma(N)=\Sigma(50)\) is held fixed.

Writing
\begin{equation}
 \Xi
 =
 \begin{pmatrix}
  \Xi_{RR} & \Xi_{RD}\\
  \Xi_{DR} & \Xi_{DD}
 \end{pmatrix},
 \qquad
 \begin{aligned}
 R&=(\phi,\tau,u_\phi,u_\tau),\\
 D&=(\chi,\sigma,u_\chi,u_\sigma),
 \end{aligned}
\end{equation}
the retained covariance is the marginal principal block
\begin{equation}
 \boxed{\Xi_{\rm ret}=\Xi_{RR}.}
 \label{eq:marginal_covariance_supp}
\end{equation}
Because the discarded variables are marginalized rather than conditioned to vanish, this is not the Schur-complement covariance
\begin{equation}
 \Xi_{RR}-\Xi_{RD}\Xi_{DD}^{-1}\Xi_{DR}.
\end{equation}
Thus the imaginary directions are removed only after solving the full coupled mode problem and assembling the complete phase-space covariance.

On the reflection-symmetric reference background, \(\Xi_{RD}\) and the discarded-to-retained block of the linearized drift vanish to numerical precision. Dedicated full-versus-reduced propagation and retained-covariance tests pass. They certify the reduction on the reference background, not the nonlinear stochastic closure tested separately below.

The reduced stochastic process is
\begin{equation}
 \dd{\mathsf X}^I
 =
 B^I({\mathsf X})\,\dd N
 +
 \Sigma^I{}_a(N)\,\dd W_N^a,
 \qquad
 \Sigma\Sigma^{\mathsf T}=\Xi_{\rm ret},
 \label{eq:reduced_SDE_supp}
\end{equation}
with
\begin{equation}
 \langle\dd W_N^a\dd W_N^b\rangle
 =
 \delta^{ab}\dd N.
\end{equation}
The drift is evaluated on each nonlinear stochastic state, while the noise covariance is inherited from the reference-background mode calculation. The additive-noise Heun update used in production is
\begin{align}
 {\mathsf X}^{\rm p}_{n+1}
 &= {\mathsf X}_n+B({\mathsf X}_n)\Delta N
   +\Sigma(N_n)\Delta W_n,
 \nonumber\\
 {\mathsf X}_{n+1}
 &= {\mathsf X}_n+\frac{\Delta N}{2}
 \left[B({\mathsf X}_n)+B({\mathsf X}^{\rm p}_{n+1})\right]
 +\Sigma(N_n)\Delta W_n,
 \label{eq:heun_supp}
\end{align}
where \(\Delta W_n\sim{\cal N}(0,\Delta N\,\mathbf1)\) and the same increment is used in predictor and corrector. Because \(\Sigma\) depends only on clock time, the It\^o and Stratonovich prescriptions coincide here. After each completed step, the loss condition is tested first; from \(N\ge38\), first entrance into either capture set is then tested at the new endpoint. No within-step hitting-time interpolation or hysteresis is applied.

Every history starts at \(N_i=20\) from the same deterministic phase-space point,
\begin{equation}
 {\mathsf X}_i
 =
 (\bar\phi,\bar\tau,\bar\phi',\bar\tau')_{N=20},
 \label{eq:prepared_supp}
\end{equation}
with no additional random initial displacement.

Before computing finite-history probabilities, we quantify local access to the competing deterministic continuation. Let \(C(N)\) be the covariance accumulated by the linearized reduced stochastic system about the reference trajectory from \(N_i\) to \(N\). Explicitly,
\begin{equation}
 \begin{aligned}
 C'&=AC+CA^{\mathsf T}+\Xi_{\rm ret},
 &C(20)&=0,\\
 A^I{}_J(N)
 &=\left.\frac{\partial B^I}{\partial{\mathsf X}^J}
 \right|_{\bar{\mathsf X}(N)} .
 \end{aligned}
 \label{eq:linear_covariance_supp}
\end{equation}
Equation~\eqref{eq:linear_covariance_supp} is integrated with DOP853 over \(20\le N\le48\), using \({\tt rtol}=2\times10^{-9}\), \({\tt atol}=2\times10^{-12}\), and maximum step \(0.03\). The drift Jacobian is evaluated by centered finite differences, and residual floating-point violations of positive semidefiniteness are projected to the nearest nonnegative spectrum. Let
\(E_\perp\) inject \((\tau,u_\tau)\) into \({\mathsf X}\). We define
\begin{equation}
 C_\perp(N)
 =
 E_\perp^{\mathsf T}C(N)E_\perp
 =
 L_\perp(N)L_\perp^{\mathsf T}(N)
 \label{eq:Cperp_supp}
\end{equation}
on the positive covariance support; \(L_\perp\) is obtained spectrally after the same nonnegative-spectrum projection.

For deterministic evolution initialized at \(N\), the terminal time is fixed to \(N_t=48\) and \(c_{N,48}[{\mathsf X}]\) is the sign of \(\tau(48)\). The covariance-whitened transverse switching distance is
\begin{equation}
 \begin{aligned}
 d_{\rm sep}^{(\perp)}(N)
 &=
 \inf_{\xi\in\mathbb R^2}
 \Bigl\{
 \|\xi\|_2:\,
 c_{N,N_t}\!\left[
 \bar{\mathsf X}(N)+E_\perp L_\perp(N)\xi
 \right]
 \\[-2pt]
 &\hspace{3.7cm}
 \ne
 c_{N,N_t}\!\left[\bar{\mathsf X}(N)\right]
 \Bigr\},
 \end{aligned}
 \label{eq:dsep_perp_supp}
\end{equation}
with
\begin{equation}
 Q_{\rm acc}^{\perp}(N)
 =
 \frac{1}{d_{\rm sep}^{(\perp)}(N)}.
 \label{eq:Qacc_perp_supp}
\end{equation}
The minimization uses nonlinear deterministic shooting in \((\tau,u_\tau)\), holding \((\phi,u_\phi)\) at their reference values at the same \(N\). We first scan \(28\le N\le45.5\) over transverse directions and radial displacements \(0\le\|\xi\|\le12\), then iteratively refine both the switching boundary and the neighborhood of the maximizing time with successively smaller deterministic Heun steps down to \(\Delta N=0.004\). In the final three-time audit about the preliminary maximum, reducing the step from \(0.006\) to \(0.004\) changes \(Q_{\rm acc}^{\perp}\) by at most \(2.13\times10^{-3}\) relatively across the three realizations. The reported \(Q_{\rm acc}^{\perp,\max}\) is the largest value surviving this refinement and is not asserted to be an analytic global maximum.

Hence \(Q_{\rm acc}^{\perp}\) is a local two-dimensional phase-space diagnostic, not a global four-dimensional distance or a first-passage probability. Its value also depends on the covariance-induced whitening convention; \(Q_{\rm acc}^{\perp}=O(1)\) means only that the local switching distance is comparable to the accumulated transverse covariance scale.

For \(\alpha=(0.8,1,1.2)\),
\begin{equation}
 Q_{\rm acc}^{\perp,\max}
 =
 (0.9092,\,1.0126,\,1.1107).
 \label{eq:Qacc_values_supp}
\end{equation}
Thus local switching becomes covariance-scale accessible during the same episode in which the tracked branch mode is temporarily light. The probability of reaching a recovered outgoing history, however, remains a finite-history quantity.

Recovered classical histories are identified only after both configuration and velocity retracking have occurred in a recovered heavy regime. We make this criterion explicit by defining, for each branch label \(s=\pm1\), the configuration- and velocity-tracking errors
\begin{align}
 \Delta_s^{(q)}
 &=
 \frac{|\tau-sv|}{v},
 \nonumber\\
 \Delta_s^{(u)}
 &=
 \frac{|u_\tau-sv_{,\phi}u_\phi|}
 {|sv_{,\phi}u_\phi|+u_{\rm reg}},
 \qquad
 u_{\rm reg}=10^{-8}.
 \label{eq:tracking_errors_supp}
\end{align}
Here \(\Delta_s^{(q)}\) measures displacement from the instantaneous branch center, while \(\Delta_s^{(u)}\) measures recovery of the velocity required to follow that moving center.

The heavy-sector recovery gate is characterized by
\begin{align}
 {\cal T}_{\rm tr}
 &=
 \frac{H^2}{m_{\tau,\rm tr}^2}
 \left|
 \ell'+\ell^2+(3-\epsilon_H)\ell
 \right|,
 \nonumber\\
 {\cal A}_m
 &=
 \left|
 \frac{H}{m_{\rm br,tr}}
 \frac{\dd\ln m_{\rm br,tr}}{\dd N}
 \right|,
 \qquad
 m_{\rm br,tr}
 =
 \sqrt{\mu_{\rm br,tr}^2},
 \label{eq:recovery_supp}
\end{align}
where
\begin{equation}
 m_{\tau,\rm tr}^2
 =
 2\gamma{\cal E}v^2 .
 \label{eq:tracking_mass_supp}
\end{equation}
The scale \(m_{\tau,\rm tr}\) controls the quasistatic positional response, whereas \(\mu_{\rm br,tr}^2\) is the continuity-tracked branch diagnostic used to establish spectral re-heaviness.

We then define the recovered-channel capture set
\begin{align}
 \Gamma_s(N)
 =
 \bigg\{
 {\mathsf X}:\;&
 s\tau>0,\quad
 \Delta_s^{(q)}<0.10,\quad
 \Delta_s^{(u)}<0.20,
 \nonumber\\
 &
 {\cal T}_{\rm tr}<0.10,\quad
 \frac{m_{\rm br,tr}}{H}>2.0,\quad
 {\cal A}_m<0.20
 \bigg\}.
 \label{eq:capture_supp}
\end{align}
Thus \(\Gamma_s(N)\) is the time-dependent region of reduced phase space in which a history has the sign of branch \(s\), has recovered both positional and velocity tracking of that branch, and satisfies the adopted heavy-sector recovery conditions. First entrance into \(\Gamma_s\), rather than crossing \(\tau=0\) or changing an instantaneous local classifier, defines recovery onto the outgoing continuation \(s\).

In the production prescription, the heavy-sector quantities \({\cal T}_{\rm tr}\), \(m_{\rm br,tr}/H\), and \({\cal A}_m\) are evaluated from the reference-background recovery profile, while \(s\tau\), \(\Delta_s^{(q)}\), and \(\Delta_s^{(u)}\) are evaluated on each stochastic trajectory. The resulting capture rule is therefore a reference-background heavy-sector gate supplemented by pathwise configuration and velocity tracking. Its branch-label sensitivity to an auxiliary trajectory-local heavy-sector prescription is tested independently below.

The distinction between spectral recovery and dynamical retracking follows directly from the response to a moving branch. Writing
\begin{equation}
 \tau(N)=s\,v[\phi(N)]+\delta\tau(N)
\end{equation}
and linearizing about the moving branch after re-heaviness gives
\begin{equation}
 \frac{\delta\tau}{sv}
 \simeq
 -\frac{H^2}{m_{\tau,\rm tr}^2}
 \left[
 \ell'+\ell^2+(3-\epsilon_H)\ell
 \right].
 \label{eq:tracking_response_supp}
\end{equation}
Hence, recovery of a heavy branch sector does not imply instantaneous recovery of the moving classical channel: positional and velocity retracking can follow after a finite relaxation interval.

The controlled numerical domain is
\begin{equation}
 {\cal D}_{\rm ctrl}
 =
 \left\{
 {\mathsf X}:
 {\mathsf X}\ {\rm finite},\
 \epsilon_H<2.8,\
 |\tau|<1
 \right\}.
 \label{eq:control_supp}
\end{equation}
The \(\epsilon_H\) cut keeps the evolution away from the singular surface \(3-\epsilon_H=0\) and is not an inflationary end condition. Crossing \(\tau=0\), changing the sign of the transverse field, or changing the local deterministic classifier does not by itself constitute history selection. Selection occurs only through first passage into one of the recovered-channel sets \(\Gamma_s(N)\).

Define the competing recovery and loss times by
\begin{equation}
 \begin{aligned}
 T_s
 &=
 \inf\{N\ge N_i:{\mathsf X}(N)\in\Gamma_s(N)\},
 \qquad s=\pm1,
 \\
 T_{\rm loss}
 &=
 \inf\{N\ge N_i:
 {\mathsf X}(N)\notin{\cal D}_{\rm ctrl}\}.
 \end{aligned}
 \label{eq:hitting_times_supp}
\end{equation}
The corresponding finite-horizon first-passage probabilities are
\begin{equation}
 h_s({\mathsf X},N)
 =
 \Prob_{{\mathsf X},N}
 \left(
 T_s<T_{-s},\
 T_s<T_{\rm loss},\
 T_s\le N_f
 \right),
 \label{eq:committor_explicit_supp}
\end{equation}
and the conditional weights of the two recovered classical continuations are
\begin{equation}
 p_s^{\rm sel}
 =
 \frac{h_s}{h_++h_-},
 \qquad
 h_++h_->0.
 \label{eq:conditional_weight_supp}
\end{equation}
For a finite ensemble, this becomes
\begin{equation}
 \boxed{
 \widehat p_s^{\rm sel}
 =
 \frac{n_s}{n_++n_-},
 }
 \label{eq:psel_estimator_supp}
\end{equation}
where \(n_s\) is the number of histories whose first recovered-channel entrance is into \(\Gamma_s\).

\begin{table}[tbp]
\centering
\caption{Local accessibility, finite-history selection, the reference-background heavy-sector gate sustained through \(N_f=54\), and subsequent recovery of reference-background tracking.}
\label{tab:selection_supp}
\begin{tabular}{@{}lccc@{}}
\toprule
 & \(\alpha=0.8\) & \(\alpha=1.0\) & \(\alpha=1.2\)\\
\midrule
\(Q_{\rm acc}^{\perp,\max}\)
& \(0.9092\) & \(1.0126\) & \(1.1107\)\\
\(\widehat p_+^{\rm sel}\)
& \(0.3343\) & \(0.3705\) & \(0.3957\)\\
\(N_{\rm heavy}\)
& \(48.48\) & \(48.65\) & \(48.77\)\\
\(N_{\rm track}\)
& \(48.90\) & \(49.08\) & \(49.24\)\\
\(N_{\rm track}-N_{\rm heavy}\)
& \(0.41\) & \(0.43\) & \(0.46\)\\
\bottomrule
\end{tabular}
\end{table}

Here \(N_{\rm heavy}\) is the first point on a uniform 500-point grid over \(38\le N\le54\) after which the three reference-background conditions \({\cal T}_{\rm tr}<0.10\), \(m_{\rm br,tr}/H>2.0\), and \({\cal A}_m<0.20\) remain satisfied at every later grid point through \(N_f=54\). The later \(N_{\rm track}\) is found analogously on a uniform 1600-point grid over \(\max(38,N_{v,\min})\le N\le54\), using the configuration-and-velocity conditions in Eq.~\eqref{eq:tracking_errors_supp}. Individual
stochastic histories recover at their pathwise times \(T_s\), not at a universal \(N_{\rm track}\). In all three realizations,
\begin{equation}
 N_{\rm heavy}<N_{\rm track},
 \label{eq:heavy_track_order_supp}
\end{equation}
showing a finite interval between horizon-persistent heavy-sector recovery and reference-background retracking. ``Persistent'' here refers only to the certified interval through \(N_f=54\), not to an extrapolation beyond the audited horizon.

For the central \(20000\)-history production ensemble,
\begin{equation}
 \widehat p_+^{\rm sel}=0.3705,
 \qquad
 0.3638<p_+^{\rm sel}<0.3772
 \label{eq:central_wilson_supp}
\end{equation}
is the corresponding \(95\%\) Wilson interval. By contrast, \(\varepsilon_q=0\) is the exact deterministic initial-value endpoint for the same prepared state; it resolves onto the negative branch and therefore
\begin{equation}
 p_+^{\rm sel}(\varepsilon_q=0)=0.
 \label{eq:zero_noise_supp}
\end{equation}

The source-deformation scan in Fig.~\ref{fig:noise_scan} uses
\begin{equation}
 \begin{aligned}
 \varepsilon_q\in\{&
 1,0.9,0.8,0.7,0.6,0.5,0.45,0.4,\\
 &0.35,0.3,0.25,0.2,0.15,0.1,0\}.
 \end{aligned}
 \label{eq:noise_grid_supp}
\end{equation}
For each realization, the physical-source point uses \(20000\) histories, each intermediate nonzero point uses \(5000\), and the zero-source point is the single deterministic initial-value solve. The nonzero ensembles use common random numbers generated from the master seed quoted below, reducing sampling noise in the source-deformation comparison. Pointwise error bars are 95\% Wilson intervals. Only \(\Sigma\) is multiplied by \(\varepsilon_q\); the prepared state, nonlinear drift, capture and loss
rules, timestep, and finite horizon are unchanged.

Every sampled history in the three primary production ensembles resolves into one outgoing branch before the horizon, with no sampled loss or unresolved endpoint. This empirical finite-ensemble result is not promoted to the exact identity \(h_++h_-=1\); the underlying finite-horizon process allows loss or unresolved probability.

As an independent recovery-definition audit, \(1500\) shared-noise histories per realization were classified simultaneously with the production reference-background gate and an auxiliary trajectory-local heavy-sector prescription. In the latter, \(H\) and the deterministic drift are evaluated at the stochastic state, \(m_{\rm br,tr}/H\) is replaced by the instantaneous lightest entropy eigenmass at that state, and \({\cal A}_m\) is obtained from a centered deterministic-flow probe of width \(\Delta N=0.002\); the same configuration and velocity thresholds are retained. Because both prescriptions act on the same stochastic histories, label differences would isolate the recovery definition rather than the noise realization. All histories resolved by both prescriptions receive the same outgoing label. Across the three realizations, \(88.6\%\)--\(89.1\%\) are captured on the same SDE step; the remaining doubly resolved histories differ only in capture time, with the trajectory-local prescription identifying recovery earlier. Thus, no branch-label sensitivity to the tested prescription is resolved at the timestep and ensemble size of this audit. This does not establish invariance under arbitrary recovery definitions.

Finally, representative captured states from both outgoing branches are restarted from their exact phase-space coordinates with
\begin{equation}
 \Sigma^I{}_a=0.
 \label{eq:noise_off_restart_supp}
\end{equation}
No projection onto the branch center, velocity reset, or matching condition is imposed. The trajectories then approach the corresponding deterministic outgoing solutions. Thus the later branch motion is sustained by the ordinary classical equations rather than by continued stochastic forcing.

\subsection{Numerical certification and stochastic-closure diagnostics}
\label{sec:controls}
The primary calculation uses a Radau background solver with \({\tt rtol}=2\times10^{-9}\), \({\tt atol}=10^{-12}\), and maximum step \(0.05\), and a DOP853 mode solver with \({\tt rtol}=2\times10^{-9}\), \({\tt atol}=2\times10^{-12}\), and maximum step \(0.04\) for the stochastic shell kernel; the normalization and pre-selector transfer runs use maximum step \(0.05\). The stochastic source uses \(220\) shell crossings at \(\rho=0.10\); each production ensemble contains \(20000\) histories evolved from \(N_i=20\) to \(N_f=54\) with \(\Delta N_{\rm SDE}=0.005\). The master seed is \(20260829\).

The principal stochastic approximation is that \(\Sigma^I{}_a(N)\) is derived from the coupled modes of the rolling reference history and used as a time-dependent source in the nonlinear ensemble. The drift remains state-dependent, but the ultraviolet mode problem and noise covariance are not recomputed along each realization. This defines the background-derived Markov closure used here. The sharp shell prescription is also Markovian; smoother coarse-graining can generate temporally correlated noise \cite{Casini:1998wr}, which lies outside the present calculation. Recomputing the complete central shell kernel at \(\rho=0.05,0.10,0.20\), with the same shell grid, componentwise spline, positive-semidefinite factorization, and endpoint hold, and evolving \(20000\) common-random-number histories per value gives \(\widehat p_+^{\rm sel}=0.3294,0.3705,0.4057\), respectively, with no loss or unresolved endpoint. Thus the nonzero competing population survives this range, while the resolved shift shows that its quantitative weight is sharp-shell-prescription sensitive.

To diagnose departures from this covariance closure, let \(\bar{\mathsf X}(N_{\rm near})\) be the nearest reference-history state to a sampled stochastic state \({\mathsf X}\) in the scaled local phase-space metric. We define
\begin{align}
 \delta_H
 &=
 \frac{|H({\mathsf X})-H(\bar{\mathsf X})|}
 {H(\bar{\mathsf X})},
 \nonumber\\
 \delta_{\cal M}
 &=
 \frac{
 \|({\cal M}/H^2)_{\mathsf X}
 -({\cal M}/H^2)_{\bar{\mathsf X}}\|_F
 }{
 \|({\cal M}/H^2)_{\bar{\mathsf X}}\|_F
 },
 \nonumber\\
 \delta_\Xi
 &=
 \frac{
 \|\Xi(N_{\rm near})-\Xi(N)\|_F
 }{
 \|\Xi(N)\|_F
 } .
 \label{eq:closure_defs_supp}
\end{align}
Here \(\delta_H\) and \(\delta_{\cal M}\) compare the stochastic state with its nearest reference-history state, while \(\delta_\Xi\) measures the covariance change between assigning the precomputed source by clock time \(N\) and by reference-history location \(N_{\rm near}\).

The recorded diagnostics show non-negligible mass-operator departures, while the Hubble and covariance-location mismatches are smaller. They quantify the departure of sampled states from the reference-background closure; they do not provide an error estimate on the branch probabilities or replace a fully state-dependent ultraviolet mode calculation.

The principal numerical tests are summarized in Table~\ref{tab:validation_supp}. Timestep comparisons use paired common-random-number ensembles, so branch disagreement directly probes pathwise discretization stability.

The recovery-profile stress test varies the thresholds in the order \((\Delta^{(q)},\Delta^{(u)},{\cal T}_{\rm tr},m_{\rm br,tr}/H,{\cal A}_m)\):
\begin{align*}
 {\rm default}&:\ (.10,.20,.10,2.0,.20),\\
 {\rm loose}&:\ (.15,.30,.15,1.7,.30),\\
 {\rm strict\ stress}&:\ (.06,.12,.06,2.5,.12).
\end{align*}
Each profile uses the same first-endpoint capture semantics as Eq.~\eqref{eq:capture_supp}, with \(3000\) histories and master seed \(20260829\).

\begin{table}[tbp]
\centering
\caption{Numerical certification of the primary calculation. Ranges span the three \(\alpha\) realizations where applicable; timestep and short-shell tests use the central realization.}
\label{tab:validation_supp}
\small
\begin{tabular}{@{}lp{0.64\columnwidth}@{}}
\toprule
Test & Result\\
\midrule
Branch equation
& \(8.39\times10^{-11}\)--\(2.63\times10^{-9}\) (scaled)\\
Curvature identity
& \(1.45\times10^{-12}\)--\(2.11\times10^{-12}\) (scaled)\\
Entropy closure
& \(4.26\times10^{-14}\) maximum absolute residual\\
Wronskian
& \(1.023\times10^{-8}\) maximum relative residual\\
SDE timestep
& \(0.005\to0.0025\to0.00125\);
  maximum paired branch disagreement \(2\times10^{-4}\)\\
Mode initialization
& \(x_{\rm init}=60,80,120\);
  covariance shift \(<3.94\times10^{-4}\)\\
Shell decimation \(\times2\)
& \(1.142\times10^{-3}\) covariance shift\\
Shell decimation \(\times4\)
& \(1.798\times10^{-2}\) covariance shift\\
Short-shell pairing
& \(2500\) histories; \(\Delta p=0\), \(n_{\rm disagree}=0\)\\
Coarse graining \(\rho\)
& \(\widehat p_+^{\rm sel}=0.3294\)--\(0.4057\); all \(60000\) resolved\\
Recovery profiles
& loose/default/strict-stress give identical branch counts\\
Capture audit
& identical branch labels for all doubly resolved histories\\
\bottomrule
\end{tabular}
\end{table}

For the central dedicated \(5000\)-history timestep ensemble,
\begin{equation}
 \widehat p_+^{\rm sel}
 =
 (0.3584,\,0.3584,\,0.3586)
\end{equation}
on the three nested grids. The maximum paired disagreement is \(2\times10^{-4}\), and all three grids have zero lost or unresolved histories. These are dedicated common-random-number certification ensembles, not alternative estimates of the \(20000\)-history production weights.

The shell, mode-initialization, and recovery tests similarly probe numerical stability within the adopted stochastic construction, not the physical validity of the background-derived covariance closure itself. Within that closure, the positive-history population is stable under all tested numerical variations, while suppressing the stochastic source returns the system continuously to the deterministic negative-branch endpoint.

For the Wronskian and mode-initialization audits, representative modes are chosen at the light-window entrance, \(N_{v,\min}\), \(N_{\mu,\min}\), and the light-window exit. The \(x_{\rm init}\) test recomputes those coupled modes at \(x_i=60,80,120\). Shell decimation retains every second or fourth shell point, including the endpoint, and reconstructs the full-grid kernel with the same componentwise cubic spline before comparing covariances. The short-shell test then evolves \(2500\) paired common-random-number histories with the full and decimated kernels. These prescriptions underlie the corresponding rows of Table~\ref{tab:validation_supp}.

\subsection{Pre-selector curvature two-point function}
\label{sec:transfer}
We separately test whether the selector episode produces a resolved change in the curvature two-point function of modes that exited beforehand, using the same complete coupled-mode system on the reference background.

With
\begin{equation}
 T^A
 =
 \frac{q^{A\prime}}
 {\sqrt{q^{B\prime}q'_B}},
\end{equation}
the adiabatic mode, curvature perturbation, and curvature power are
\begin{align}
 Q_{\parallel,k\lambda}
 &=
 T_AQ^A_{k\lambda},
 \nonumber\\
 {\cal R}_{k\lambda}
 &=
 \frac{Q_{\parallel,k\lambda}}
 {\sqrt{2\epsilon_H}},
 \label{eq:adiabatic_projection_supp}\\
 {\cal P}_{\cal R}(k,N)
 &=
 \frac{k^3}{2\pi^2}
 \sum_{\lambda=1}^{4}
 |{\cal R}_{k\lambda}(N)|^2 .
 \label{eq:curvature_power_supp}
\end{align}
These definitions also enter the scalar-amplitude normalization in Eq.~\eqref{eq:cmb_norm_supp}.

For pre-selector modes we compare the same coupled-mode solution at \(N_{\rm pre}=33\) and \(N_{\rm post}=53\), defining
\begin{equation}
 {\cal C}_{\rm tr}(k)
 =
 \frac{
 {\cal P}_{\cal R}(k,N_{\rm post})
 -
 {\cal P}_{\cal R}(k,N_{\rm pre})
 }{
 {\cal P}_{\cal R}(k,N_{\rm pre})
 } .
 \label{eq:Ctr_supp}
\end{equation}
We test the seven equally spaced internal horizon-exit coordinates
\begin{equation}
 N_k=5+\frac{7j}{6},
 \qquad j=0,\ldots,6,
 \label{eq:transfer_mode_grid_supp}
\end{equation}
where
\begin{equation}
 k=a_{\rm sc}(N_k)H(N_k).
\end{equation}
Each is a single coupled-mode solution initialized at \(x_i=80\) and evolved with DOP853 maximum step \(0.05\); the powers at \(N_{\rm pre}=33\) and \(N_{\rm post}=53\) are read from that same solution. No reheating-dependent identification with a fixed present-day \({\rm Mpc}^{-1}\) interval is made.

With the conservative numerical detection threshold \(\delta_{\rm tr}^{\rm num}=10^{-8}\), all tested modes satisfy
\begin{equation}
 \boxed{
 |{\cal C}_{\rm tr}(k)|<10^{-8}.}
 \label{eq:transfer_sensitivity_supp}
\end{equation}
The converged numerical differences are of order \(10^{-11}\), but because \({\cal C}_{\rm tr}\) subtracts nearly equal powers, these smaller values are not interpreted as a physical upper bound. The robust result is therefore a numerical non-detection at \(10^{-8}\) sensitivity for the tested pre-selector modes on the reference background.

This test concerns only the ordinary curvature two-point function propagated by the linear coupled-mode system on the reference background. It is not a branch-conditioned late-time power spectrum of the nonlinear stochastic ensemble and does not probe observables generated by later branch-dependent physics.

Within this scope, the non-detection coexists with an order-\(10^{-1}\) redistribution among the outgoing histories. The calculation therefore distinguishes local pre-selector curvature power from the finite-history probabilities with which different classical continuations are ultimately occupied.

\bibliography{references}

\end{document}